%% file: paper.tex
\documentclass[copyright]{eptcs}
\input{macros}

\title{Approaching Symbolic Parallelization by\\Synthesis of Recurrence Decompositions}
\author{Grigory Fedyukovich and Rastislav Bod\'{i}k
\institute{Computer Science and Engineering, University of Washington, Seattle, Washington, USA}}

\def\titlerunning{Symbolic Parallelization}
\def\authorrunning{G. Fedyukovich \& R. Bod\'{i}k}

\begin{document}

\maketitle

\input{abstract}
\input{intro}
\input{language}

\input{algorithm}
\input{evaluation}
\input{related}

\bibliographystyle{eptcs}
\bibliography{references}
\end{document}

%% file: macros.tex
\usepackage{amsmath}
\usepackage{amssymb}
\usepackage{amsthm}
\usepackage[referable]{threeparttablex}
\usepackage{graphicx}
\usepackage{color}
\usepackage{xspace}
\usepackage{booktabs}
\usepackage{xhfill}
\usepackage[font=footnotesize]{caption}
\usepackage{algorithmicx}
\usepackage[tworuled,linesnumbered,algo2e]{algorithm2e}
\usepackage{fancyvrb}
\usepackage[euler]{textgreek}

\let\textquotedbl="

\def\ExclamationPoint{\char59}
\catcode`;=\active

\RecustomVerbatimCommand{\VerbatimInput}{VerbatimInput}%
{ 
 defineactive=\def;{\color{gray}\ExclamationPoint},
 framerule=0.25mm,
 numbers=left,
 frame=lines,  
 framesep=2em, 
 labelposition=topline,
}

\newcommand{\rosette}{\textsc{Rosette}\xspace}
\newcommand{\sympa}{\textsc{GraSSP}\xspace}

\newcommand{\inp}{\mathit{input}}
\newcommand{\Inp}{\mathit{In}}
\newcommand{\inpu}{\mathit{input}}
\newcommand{\outpu}{\mathit{output}}

\newcommand{\Out}{\mathit{Out}}
\newcommand{\fold}{\mathit{fold}}

\newcommand{\length}{\mathit{length}}
\newcommand{\append}{\mathit{append}}

\newcommand{\agg}{\mathit{merge}}
\newcommand{\comp}{\mathit{merge}}
\newcommand{\refine}{\mathit{completed}}
\newcommand{\cond}{\mathit{cond}}
\newcommand{\prefix}{\mathit{prefix}}

\newcommand{\cmnt}[1]{\Comment{{ #1 }}}

\newcommand{\tuple}[1]{\langle #1 \rangle} 

\newcommand{\ttimes}{\times \ldots \times}

\definecolor{midgrey}{rgb}{0.3,0.3,0.3}
\definecolor{darkred}{rgb}{0.7,0.1,0.1}
\definecolor{darkgreen}{RGB}{33,138,28}

%% file: abstract.tex
\begin{abstract}

We present \sympa, a novel approach to perform automated parallelization relying on recent advances in formal verification and synthesis.
\sympa augments an existing sequential program with an additional functionality to decompose data dependencies in loop iterations, to compute partial results, and  to compose them together.
We show that for some classes of the sequential prefix sum problems, such parallelization can be performed efficiently.

\end{abstract}

%% file: intro.tex
\section{Introduction}
\label{sec:intr}

Parallelization of software is important for improving its effectiveness and productivity.
Industrial applications (studied, e.g., in~\cite{DBLP:conf/sosp/RaychevMM15}) operate with large inputs and therefore could run days.
%
But even a small program that iterates over a single array and incrementally computes a single numerical output could be challenging for parallelization.
It would require partitioning of the input data into a sequence of segments, processing each segment separately, and merging the partial outputs for the segments.

Data dependencies prevent parallel processing of the segments.
Therefore, the parallel loop should break the dependencies by devising an alternative function. 
This additional processing could invoke some analysis either \emph{before} the parallel execution: in order to move segment boundaries,
 or \emph{after} the parallel execution: in order to repair the outputs broken by violated dependencies.
In this paper, we address the challenge of synthesizing such additional code automatically.

Preserving equivalence is a crucial requirement to automatic parallelization, and
it aims at confirming that pairwise equivalence of inputs implies pairwise equivalence of outputs
~\cite{KK1997,DBLP:conf/oopsla/0001SCA13,DBLP:conf/dac/ClarkeKY03,os2009,DBLP:conf/cav/LahiriHKR12,pde}.
%
Rather than merely relying on the existing solutions to \emph{verify} equivalence between programs, we aim to \emph{synthesize} a parallel program that is equivalent to the given sequential program.
Given arbitrary segments of the input data, we build on an assumption that segment boundaries can always be adjusted, and the actual loop computation  can be performed for the updated segments.

Program synthesis is an approach to generate a program implementation from the given specification.
State-of-the-art synthesis tools employ the Counter-Example-Guided Inductive Synthesis (CEGIS)~\cite{DBLP:conf/asplos/Solar-LezamaTBSS06} paradigm, i.e.,  assume a space of candidate implementations and check whether there exists a candidate among them that matches the given specification.
%
In this work, we exploit the capabilities of SMT-based bounded model checking and template-based synthesis in \rosette~\cite{DBLP:conf/oopsla/TorlakB13,DBLP:conf/pldi/TorlakB14}.

Our novel approach \sympa automatically generates a search space of
candidate decompositions for the given program, each of which consists of function $\prefix$ (that would identify during runtime
how segment boundaries should be moved),
and function $\comp$ (that would identify how the partial outputs should be combined to get the final output).
Then, for a given number of segments and for a given maximal length of each segment, \sympa chooses a candidate decomposition and formally verifies if it produces results equivalent to the given sequential program.
Our preliminary experiments with \sympa confirm that for reasonably small programs the decomposition can be found in seconds.

%% file: language.tex
\section{Sequential Recurrence Decomposition}
\label{sec:pre}

We start by introducing a functional notation for array-handling programs, and proceed by formulating the parallelization criteria for them.
We focus on functions taking a single finite-sized array as input, and recurrently computing a single output.

An $\inpu$ and an $\outpu$ are the designated variables respectively of  type $\Inp$ and $\Out$.
%
%
An $n$-sized \emph{array} is a finite sequence of inputs:%
\begin {equation}\notag
A : \Inp^n
\end{equation}

In this paper, we consider functions of the following  type:%
\begin {equation}\notag
f: D \times \Inp \rightarrow D
\end{equation}

where $D$ is a domain of any type.
An element $d$ of type $D$ is called \emph{state}.
Intuitively, $f$ takes a state and updates it with respect to a given input.
The state calculated by function $f$ can further be taken by function $h$ to compute an $\outpu$:%
\begin {align}\notag
& h: D  \rightarrow \Out 
& h(f(d,\inpu)) = \outpu
\end{align}

In the scope of the paper, we are interested in iterative application of function $f$ to the elements of an $n$-sized array by means of
 the higher-order function $\fold$:%
\begin {equation}\notag
\fold:(D \times \Inp \rightarrow D) \times D \times \Inp^n \rightarrow D
\end{equation}

For example, consider  array $A = \tuple{\inp_1,\ldots,\inp_n}$.
In the first iteration, the initial state $d_0$ is taken by $f$ and updated with respect to the first element $\inp_1$.
Then, the updated state is taken by $f$ again and updated with respect to the second element $\inp_2$.
A state obtained after $n$ iterative applications of $f$ to $d_0$ is called \emph{final}.
It is represented as the following first-order \emph{recurrent relation}:%
\begin {equation}\notag
\fold(f, d_0, A) = f(\inp_n, f(\inp_{n-1}, \ldots, f(\inp_2, f(\inp_1, d_0))))
\end{equation}

Notably, $\outpu$ has to be computed only for the final state:
\begin {equation}\notag
h(\fold(f, d_0, A)) = \outpu
\end{equation}

Throughout the paper, we rely on an operator that concatenates $m$ arrays:%
\begin {equation}\notag
\append : \Inp^{n_1} \ttimes \Inp^{n_m} \rightarrow \Inp^{n_1 + \ldots + n_m}
\end{equation}

It is important to ensure the following functional property of $\append$:%
\begin {equation}
\label{eq:seq}
\fold(f, d_0, \append(A_1, \ldots A_m)) =
   \fold(f, \fold(f, \ldots \fold(f, d_0, A_1), \ldots, A_{m-1}),  A_{m})
\end{equation}

The left-hand-side of  equation~\eqref{eq:seq} denotes the initial state $d_0$ iteratively updated by $f$ with respect to all elements of $\append(A_1, \ldots A_m)$.
The right-hand-side of the equation consists of $m$ groups of consequent applications of $\fold$ to  each of the arrays $\{A_i\}$.
That is, the final state obtained for an $i$-th application of $\fold$ is further used for the $(i+1)$-th application of $\fold$.
We refer to this property to as \emph{sequential recurrence decomposition}, since it guarantees equivalence between a single application of $\fold$ to $\append(A_1, \ldots A_m)$ and $m$ recurrent applications of $\fold$.

Trivially, the equivalence of final states entails equivalence of the outputs computed out of these states.

\section{Parallel Recurrence Decomposition}
\label{sec:par}

Application of $\fold$ to each of the arrays $\{A_i\}$ in parallel requires the recurrent relation to be decomposed.
We assume that each application of $\fold$ to $A_{i}$ takes the same initial state $d_0$,
 and refer to the corresponding final states $\{d_i\}$  as to \emph{partial} states
(and to the corresponding outputs as to \emph{partial} outputs).%
\begin{align}\notag
&\forall i \cdot d_i \triangleq \fold(f, d_0, A_i) 
&\forall i \cdot \outpu_i \triangleq  h(d_i)  
\end{align}

The question is how to find a suitable function $\agg$ that takes all partial results as input ant produces an output that is equivalent to the output of the corresponding  sequential computation:%
\begin {equation}\notag
\agg : \Out^m \rightarrow \Out
\end{equation}

%
%
%
In the rest of the section, we aim at establishing a property of  \emph{parallel} (as opposed to sequential) \emph{recurrence decomposition} that binds together all ingredients of the parallel processing of $m$ arrays.
Interestingly, there are several possible ways of defining this property, depending on existence of function $\agg$ for each particular $f$. 
We consider three such cases.


\subsection{Direct Decomposition}
\label{sec:no}

The first case assumes existence of function $\agg$ that is \emph{directly} applicable to all partial outputs obtained after applications of $\fold$ to each $A_i$:%
%
\begin {equation}
\label{eq:noref}
h(\fold(f,d_0,\append(A_1, \ldots A_m))) = \agg (\outpu_1 \ldots, \outpu_m)
\end{equation}

\begin{figure}[t!]
\begin{Verbatim}
; given functions:
    (define (f element current-max) (max element current-max))
    (define (array-max A) (foldl f -inf.0 A))

; function needed for parallel decomposition:
    (define (merge-max out) (array-max out))
\end{Verbatim}
\caption{Calculating a maximal element of the array and a suitable function $\agg$ in \rosette.}%
\label{ex:1}
\end{figure}

\paragraph{Example.} %
Consider function \texttt{array-max} that calculates a maximal element of a given array in \rosette%
\footnote{We provide $\fold$-like implementation to comply with the chosen formalization. An alternative implementation using \texttt{apply} is also possible (similar to \texttt{max} in Fig.~\ref{ex:3}).}
 (shown in Fig.~\ref{ex:1}).
Function \texttt{f} is essentially a wrapper for the built-in function \texttt{max}, i.e., it outputs a maximal element of pair $\tuple{\texttt{element}, \texttt{current-max}}$.
Function \texttt{array-max} delegates its functionality to \texttt{foldl} (which stands for left-fold) that iteratively applies \texttt{f} to the initial state $-\infty$ and to the elements of array \texttt{A}.
For some partitioning \texttt{A = (append A$_1 \ldots $ A$_m$)}, individual applications of \texttt{array-max} to each segment produce array \texttt{out}  consisting of $m$ elements, where $\texttt{out}_i$ is maximal element for \texttt{A$_i$}. 
It is easy to see that the maximal element of \texttt{A} is the maximal element of \texttt{out}.
So for \texttt{array-max} there exists function $\agg$, and it is equal to \texttt{array-max}. 
\qed

We return to this scheme in Sect.~\ref{sec:alg} and show how function $\agg$ can be synthesized.

\subsection{Decomposition with Constant Prefixes}
\label{sec:const}

When no  function $\agg$ meeting the condition~\eqref{eq:noref} exists, then the computations $\fold(f, d, A_{i})$ and $\fold(f, d, A_{i+1})$ depend on each other and cannot be \emph{correctly} performed from the speculative initial state $d=d_0$. 
%
If we are in the lucky situation that the impact of the speculative initial state is localized to a prefix of $A_i$ then we can repair the incorrect execution by recomputing the affected prefix. 
In this subsection, we show how such a repair can be performed on a constant-size prefix, if one exists.
First, computations $\fold(f, d, A_i)$ are performed in parallel from the initial state $d=d_0$, computing the partial result $d_i$. 
Subsequently, the computations are repeated for a constant-size prefix of $A_{i+1}$, starting from the state $d_i$.
The result of processing the prefix is then supplied instead of $d_i$ to a suitable  function $\agg$.

We say that $A'$ is a \emph{prefix} of $A$ if $A = \append(A', A'')$ for some $A''$. 
We allow the prefix to be an empty array.
We denote by $\prefix(A)$ the prefix of $A$, and by $\prefix_\length$ its predetermined length.
We call such prefix the \emph{constant prefix}. 

Assuming the existence of constant prefixes for each $A_{i+1}$, the corresponding partial outputs are computed and repaired as follows:%
\begin {align}\notag
&\forall i \cdot d_i^\refine \triangleq \fold(f,d_i, \prefix(A_{i+1}))
&\forall i \cdot \outpu_i^\refine \triangleq h(d_i^\refine)
\end{align}
%
%
Assuming the existence of a suitable function $\agg$, the repaired partial outputs are combined:
\begin {equation}\label{eq:simpref}
h(\fold(f,d_0,\append(A_1, \ldots A_m))) = \agg (\outpu_1^\refine \,\,, \ldots,\,\, \outpu_{m-1}^\refine, \outpu_m )
\end{equation}

Note that the partial output produced for the last array $A_m$ is always completed since no subsequent array needs to be repaired. 

An important observation is that computing each $d_i^\refine$ requires processing $\prefix(A_{i+1})$ twice, with the second processing serialized after $d_i$ has been computed. 
The inefficiency is mitigated by the observation that the prefixes can be processed in parallel, so the critical path of the computation grows only by the processing of the constant prefix. 

This scheme is applicable when there exists a constant prefix of each $A_{i+1}$ that limits the scope of the necessary recomputation.  Crucially, the prefix must not cover the whole $A_{i+1}$ or else the repair of $\fold(f, d_i, \prefix(A_{i+1}))$ would modify $d_{i+1}$, which would in turn necessitate the repair of $\fold(f, d_{i+1}, A_{i+2})$, serializing the repairs. 

\begin{figure}[t!]
\begin{Verbatim}
; given functions:
    (define (f element state)
      (cond
        [(>= element (first state)) '(element (second state))]
        [else '(0 0)]
      )
    )
    (define (is-sorted A) (foldl f '(-inf.0 1) l))

; functions needed for parallel decomposition:
    (define (merge-sorted out) (apply min out))
    (define (prefix-length-sorted) 1)
\end{Verbatim}
\caption{Checking if array is sorted, a suitable function $\agg$ and a suitable value of $\prefix\_length$.}%
\label{ex:2}
\end{figure}

\paragraph{Example.} %
Consider function \texttt{is-sorted} (shown in Fig.~\ref{ex:2}) that returns 1 if for each pair of consequent elements of array \texttt{A}, the former is smaller than the latter, and returns 0 otherwise.
Function \texttt{f} takes \texttt{element} and pair \texttt{state} whose first element gets compared to \texttt{element}, and whose second element is either passed to output (if \texttt{element} is greater) or dropped to zero (otherwise).

The merge function is  \texttt{min}: it takes  array \texttt{out} of zeroes and ones, and returns 1 if all the arrays from \{\texttt{A$_i$}\} are sorted (i.e., all elements of \texttt{out} are ones), and 0  if at least one array is not sorted (i.e., at least one element of \texttt{out} is zero).
Each individual result is necessary but not sufficient for concluding that \texttt{(append A$_1 \ldots $ A$_m$)} is sorted.
It remains to establish that all elements of each \texttt{A$_{i}$} are smaller than any element of \texttt{A$_{i+1}$}, and we do it via constant prefixes.

The prefix of each \texttt{A$_{i+1}$} is used for repairing \texttt{out$_i$} and it is defined as the array containing only the first element%
\footnote{Following the \rosette notation, \texttt{(take A$_{i+1}$ 1)} stands for the sub-array of \texttt{A$_{i+1}$} containing just its first element.}
of \texttt{A$_{i+1}$}.
Note that the prefixes should be processed twice, e.g., for checking that \texttt{(append A$_i$ (take A$_{i+1}$ 1))}
is sorted and for checking that \texttt{A$_{i+1}$} itself is sorted.%
\qed

We return to this scheme in Sect.~\ref{sec:alg} and show how function $\agg$ as well as constant $\prefix_\length$ can be synthesized.

%
%

\subsection{Decomposition with Conditional Prefixes}
\label{sec:cond}

When there is no function $\agg$ meeting condition~\eqref{eq:simpref} for all possible constant prefixes, then there could exist more complicated functions to express the length of prefixes.
Indeed, the number of elements in  $\prefix(A_{i})$
might depend on the elements themselves.
The problem of finding such \emph{conditional} prefixes can be reduced to iterative evaluation of  predicate $\prefix_\cond$ for each element of the given array:%
\begin {equation}\notag
\prefix_\cond: \Inp \rightarrow bool
\end{equation}

That is, for each $i$, the length of $\prefix(A_i)$ is the position number $k$ of some element in $A_i$, such that $\prefix_\cond$ evaluates to $\mathit{true}$ for the $k$-th element, and $\prefix_\cond$ evaluates to $\mathit{false}$ for all $j$-th elements in $A_i$ where $j < k$.
Thus, contrary to constant prefixes that could be identified by some static analysis of $f$, the calculation of conditional prefixes requires running $f$ for the particular given arrays.
For the tasks enjoying  existence of a suitable predicate $\prefix_\cond$ and existence of a suitable function $\agg$, the parallel recurrence decomposition property has the same form as~\eqref{eq:simpref}.
The difference is the way of computing prefixes, whose length is not constant any longer.

\begin{figure}[t!]
\begin{Verbatim}
; given functions:
    (define (f element state)
      (define seen-one (first state))
      (define seen-two (second state))
      (cond
        [(= element 1) '(1 seen-two)]
        [(= element 2) (if (= seen-one 1) '(seen-one 1) '(0 0))]
        [else '(seen-one seen-two)]
      )
    )
    (define (seen-2-after-1 A) (foldl f '(0 0) A))

; functions needed for parallel decomposition:
    (define (merge-search out) (apply max out))
    (define (prefix-cond-search element) (= element 2))
\end{Verbatim}
\caption{Searching if ``2'' appeared in the array some time after ``1'' and the possible implementation of $\agg$ and $\prefix$.}%
\label{ex:3}
\end{figure}

\paragraph{Example.} %
Consider function \texttt{seen-2-after-1} that checks whether ``2'' appeared in the given array some time after ``1'' (shown in Fig.~\ref{ex:3}).
Function \texttt{f} updates two flags indicating whether ``1'' or ``2'' has been already seen in the array.

The merge function is  \texttt{max}: it takes the array \texttt{out} of zeroes and ones, and returns 1 if at least one array has ``2'' appeared some time after ``1''.
The prefix of each \texttt{A$_{i+1}$} contains all elements from the beginning of \texttt{A$_{i+1}$} until the first appearance of ``2''.
Indeed, consider a case when $m=2$, \texttt{A$_1$} contains ``1'', but does not contain ``2'', and  \texttt{A$_2$} contains ``2'', but does not contain ``1''.
In this case, it is important to keep searching ``2'' in \texttt{A$_2$} (i.e., traverse all elements until ``2'' is found.
In contrast,  processing \texttt{A$_1$} and \texttt{A$_2$} solely only lead to incorrect outputs.%
\qed

We return to this scheme in Sect.~\ref{sec:alg} and show how function $\agg$ and predicate $\prefix_\cond$ can be synthesized.

%% file: algorithm.tex
\begin{figure}[t!]
  \centering
  \includegraphics[scale=0.99]{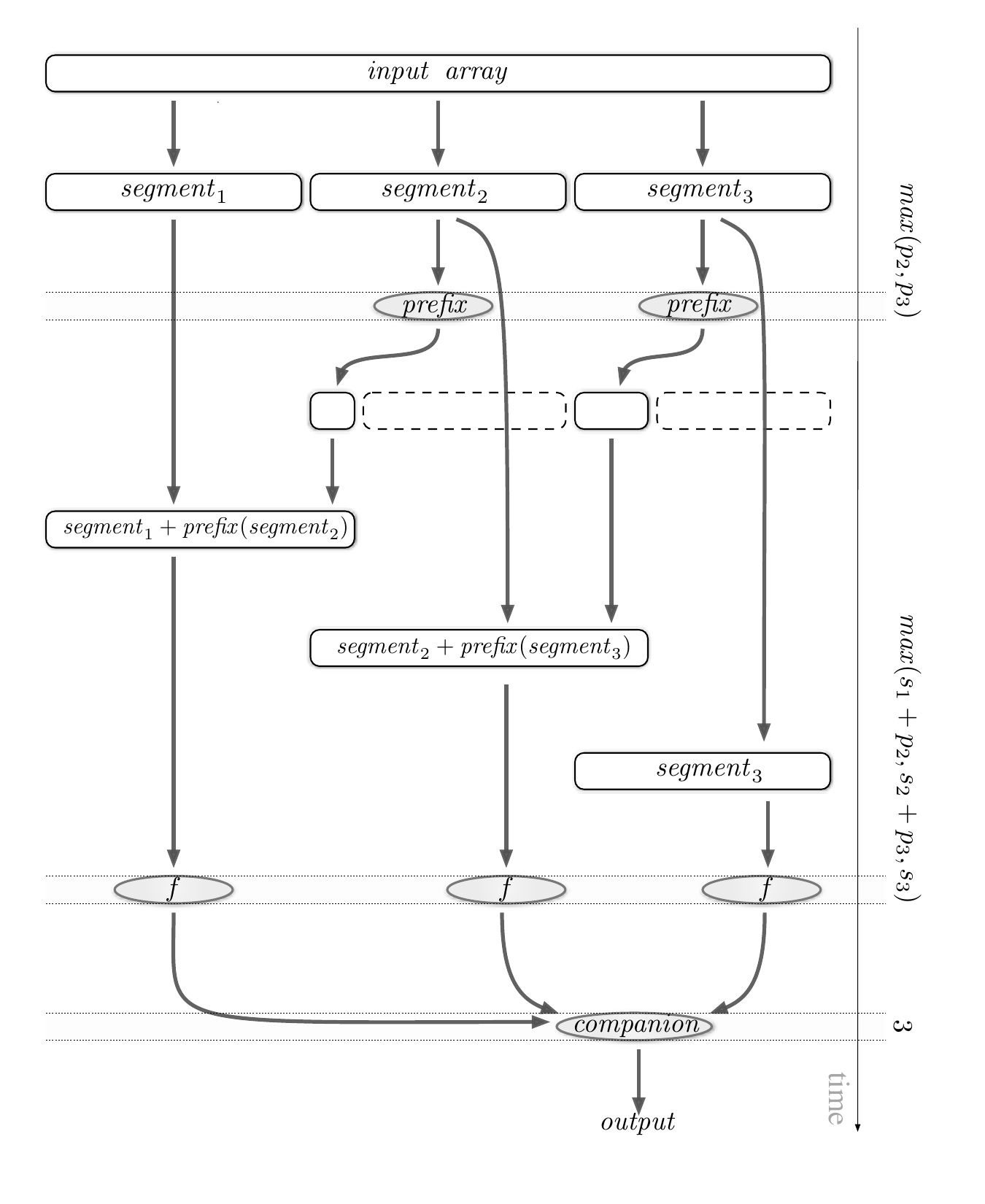}
  \caption{Executing function $f$ in parallel.}
  \label{fig:flow}
\end{figure}

\section{Bringing It All Together}
\label{sec:motiv}
Assuming the existence of function $\agg$ and (if applicable) constant $\prefix_\length$ or predicate $\prefix_\cond$ for some function $f$, we show how $f$ can be parallelized in 
Fig.~\ref{fig:flow}.
The diagram considers three processors and represents different segments of the given data (original and updated with the prefixes) with rectangular boxes, and functions to iterate over data with ovals.

We estimate the time spent at different stages of the parallel algorithm as a means of number of $\fold$-iterations.
Due to~\eqref{eq:seq}, the total time $T_s$ for three sequential applications of $\fold$ (denoted respectively $s_1$, $s_2$ and $s_3$) performed for three segments is:%
\begin{equation}\notag
T_s = s_1+s_2+s_3
\end{equation}

For the second and the third arrays, the calculation of the corresponding prefixes requires respectively times $p_2$ and $p_3$.
Since in the worst case the prefix calculation is an iterative process, $p_i$ for some $i$ might be equal to $s_i$.%
\footnote{There could also be the case when a prefix spans over the entire segment, so its computation continues for the subsequent segments. We do not consider this scenario in this paper.}
Note that  every process should wait until prefixes for all arrays are computed, and the total time $T_p$ is:%
\begin{equation}\notag
T_p = \mathit{max}(p_2,p_3)
\end{equation}

Then, for each array (already coupled with the subsequent prefix), the original function $f$ should be iteratively applied.
Here, every process should also wait for all other processes, and the total time $T_f$ is:%
\begin{equation}\notag
T_f = \mathit{max}(s_1+p_2,s_2+p_3,s_3)
\end{equation}

The last step for executing $\comp$ (called \emph{companion} in the diagram) takes time 3, since it just aggregates three integers:%
\begin{equation}\notag
T_c = 3
\end{equation}

Finally, the speedup earned by the parallel version of the function compared to the sequential one can be calculated by the following formula:%
\begin{equation}\notag
X \triangleq \frac{T_s}{T_p+T_f+T_c}
\end{equation}

\section{Parallelization Synthesis Problem}
\label{sec:alg}

Given function $f$ and initial state $d_0$, as declared in Sect.~\ref{sec:pre},
 we wish to find such function implementations for $\prefix$ and $\agg$, as declared in Sect.~\ref{sec:par}, so 
for any possible sequence of input arrays, $\prefix$ and $\agg$ correctly decompose recurrence relations foisted by $f$: %
\begin {align}\notag
\exists\, \prefix ,\agg \cdot \forall\, A_1, \ldots A_m \cdot &\\ 
h(\fold(f,d_0,\append(&A_1, \ldots A_m))) =\notag\\
\agg ( &h(\fold(f,d_0, \append(A_0, \prefix(A_{1})))),\ldots, \notag\\
 &h(\fold(f,d_0, \append(A_{m-1}, \prefix(A_{m})))),  h(\fold(f,d_0, A_{m})))\notag
\end{align}

\subsection{Our Solutions}

We present \sympa, an algorithm to deliver solutions to the synthesis problem, and outline its pseudocode in Alg.~\ref{alg:sympa}.
\sympa aims at parallelizing iterative applications of $f$ to $d_0$ for any possible input arrays $A_1,\ldots,A_m$.
It treats each $A_i$ nondeterministically, by allowing them to contain only symbolic elements.
Thus, while parallelizing $\fold(f,d_0,\append(A_1, \ldots A_m))$, the algorithm considers all possible resolution of nondeterminism in each $A_i$ and if a solution is found, it is guaranteed to be general enough to satisfy all arrays containing concrete (i.e., non-symbolic) elements.
To ensure the finiteness of the search space of the solutions, the lengths of $A_i$ are bounded.

\sympa exploits our observations made in Sect.~\ref{sec:par}, and gradually attempts synthesizing functions $\agg$ and $\prefix$ for $f$ and $d_0$ under the following hypotheses:

\begin{itemize}
\item [1)] there exists function $\agg$ that makes~\eqref{eq:noref} hold, so there is no need for prefixes (method \textsc{SyntNoPrefix}),
\item [2)] there exist function $\agg$ and function $\prefix$ with a fixed length $\prefix_\length$ that make~\eqref{eq:simpref} hold (method \textsc{SyntConstantPrefix}),
\item [3)] there exist function $\agg$ and function $\prefix$ with predicate $\prefix_\cond$ that make~\eqref{eq:simpref} hold (method \textsc{SyntConditionalPrefix}).
\end{itemize}

Each of the three methods verifies whether the corresponding hypothesis is true.
In particular, it traverses the search space of candidate implementations of $\agg$ (and, for the hypotheses 2 and 3, $\prefix$) function and checks whether one of those candidates is a witness for the hypothesis.
The increasing complexity of the methods allows saving time while parallelization, and delivering simple solutions first.
To further improve efficiency, \sympa bounds the search space of $\agg$ to contain relatively simple operators (e.g., ``+'', ``$min$'', ``$max$''), and the search space of $\prefix_\cond$ to contain conjunctions of simple terms in linear arithmetic with equality.
Notably, the efficiency comes with a price:
the more restrictions are applied to the search space, the more risks are taken by the algorithm to produce ``unknown'' output.

Given solutions to the synthesis problem, it is straightforward to compile the synthesized functions into a self-contained parallelized function that is equivalent to the sequential one, but behaves in a fashion described in Sect.~\ref{sec:motiv}.

\begin{algorithm2e}[t!]
  \SetAlgoSkip{}
  \SetAlgoNoLine
  \BlankLine
  \KwIn{Function $f$, initial state $d_0$}
  \KwOut{$\agg, [\prefix]$}
  \BlankLine
  $A_1,\ldots,A_m \gets \textsc{Nondet()}$\\
  Try \Return $ \textsc{SyntNoPrefix}(f,d_0,A_1,\ldots,A_m)$\cmnt{see Sect.~\ref{sec:no}}\\
  Try \Return $\textsc{SyntConstantPrefix}(f,d_0,A_1,\ldots,A_m)$\cmnt{see Sect.~\ref{sec:const}}\\
  Try \Return $\textsc{SyntConditionalPrefix}(f,d_0,A_1,\ldots,A_m)$\cmnt{see Sect.~\ref{sec:cond}}\\
  \Return ``unknown''
    \BlankLine
  \caption{\sympa($f, d_0$)}
  \label{alg:sympa}
\end{algorithm2e}

\subsection{Current Limitations and Future Improvements}

Since  prefixes are the key to identify the amount of overhead of  parallel computation against  sequential computation, \sympa can be required to minimize the length of discovered prefixes.
In case of constant prefixes, the algorithm can enumerate positive numbers starting from 0 and check whether the current number constitutes a sufficient prefix length for all arrays.

Interestingly, in case of conditional prefixes, it is not obvious how to construct the minimal prefix.
Since depending on evaluation of $\prefix_{cond}$, the length of each prefix can vary from array to array.
Furthermore, $\prefix_{cond}$ can be evaluated to $\mathit{false}$ for all iterations of the $i$-th application of $\fold$, implying that the entire  $A_{i-1}$ and $A_{i}$ should be concatenated, and construction of prefixes should be continued with $A_{i+1}$.
In general, the existence of $\prefix_{cond}$ does not even guarantee that a prefix can eventually be found.
And in the worst case, the entire input array should be processed sequentially.

As we mentioned in Sect.~\ref{sec:const}, for each $i$, the elements of $\prefix(A_{i})$ are processed twice: for completing the $(i-1)$-th, and for starting the $i$-th applications of $\fold$.
For the big picture, we should also mention the routine for calculating $\prefix(A_{i})$ itself that requires yet another iterative run over the elements of $A_i$.
To avoid such a redundancy,  the prefix calculation engine can be augmented by construction of symbolic summaries that would capture the effect of applying $f$ to the given prefix and an arbitrary initial state.
A similar idea was presented in~\cite{DBLP:conf/sosp/RaychevMM15}, but \sympa could enable its application in a new context.
That is, as opposed to summarizing the effect of application of $\fold$ to the entire arrays, we would like to summarize the effect of application of $\fold$ just to the prefixes, and to leave the remaining elements untouched.
Then, the computed summaries could be applied directly 1) to repair the output of the $(i-1)$-th application of $\fold$, and 2) to start the $i$-th application of $\fold$, thus avoiding duplicate iterations.
We believe, such summarization-based approach to automatic parallelization is also suitable for the cases when a prefix spans over the entire arrays(s).

%% file: evaluation.tex
\section{Evaluation}
\label{sec:eval}

We implemented a prototype of \sympa in \rosette.
Given a specification in a form of array-handling sequential function $f$, \rosette   maintains a space of candidate decompositions of $f$ and verifies equivalence between $f$ and each candidate separately.
A candidate decomposition that fulfils the specification is returned by \rosette as output.

We evaluated \sympa on a set of \rosette implementations of some prefix sum problems.
We used a bound of 15 for the length of the input array.
Interestingly, this bound was sufficient, and gave correct decompositions also for bigger arrays.
In general, of course, the soundness of recurrence decomposition for big arrays is not guaranteed.

Table~\ref{tab:eval} summarizes our preliminary results of \sympa for several benchmarks.
It measures the complexity of each benchmark by giving the number of variables in the state of each $f$ (``\# Vars'').
The results of recurrence decomposition are reported by the hypothesis which was true for the function, the operator for a synthesized function $\agg$, and a synthesized function that identifies the length of the prefix (i.e., $\prefix_\length$ for the constant prefixes, and $\prefix_\cond$ for the conditional prefixes).
Finally, we provide the total time spent by \rosette for realizability check of the correspondent hypothesis (i.e., to realize the necessity of either constant or conditional prefixes), and for synthesis of the witnessing functions.

Functions \texttt{array-max}, \texttt{is-sorted}  and \texttt{seen-2-after-1} were already discussed in Examples~\ref{ex:1},~\ref{ex:2}, and~\ref{ex:3} respectively.
Function \texttt{array-count} calculates the size of the  array.
Functions \texttt{alternation-of-} \texttt{1-2} and \texttt{alternation-of-11-22} check if the entire array if an alternation of ``1'' and ``2'' and ``11'' and ``22'' respectively.
Function \texttt{number-of-123} searches for appearances of word ``123''.
Interestingly, the synthesis time for all decompositions was insignificant.
In future work, we plan to enhance the set of experiments by more challenging benchmarks.

\begin{table}[t!]
\begin{center}
\resizebox{\linewidth}{!}{%

\begin{tabular}{lccccc}
\toprule

Benchmark 				& \# Vars 	& Hypothesis 			&  $\agg$ 		& $\prefix_\length$ / $\prefix_\cond$ & Synt time (\emph{s.})   \\ 
\midrule

\texttt{array-count} 		& 1 		& \textsc{SyntNoPrefix} 	& \texttt{+} 	& --- 					& 3 \\
\texttt{array-max} 			& 1 		& \textsc{SyntNoPrefix} 	& \texttt{max} 	& --- 					& 3 \\
\texttt{is-sorted}      			& 1	         & \textsc{SyntConstPrefix}& \texttt{min} 	& \texttt{1}				& 3 \\
\texttt{alternation-of-1-2}       	& 1	         & \textsc{SyntConstPrefix}& \texttt{min} 	& \texttt{1}				& 2 \\
\texttt{number-of-123}		& 2		& \textsc{SyntConstPrefix}	& \texttt{+}		& \texttt{2}				& 3 \\
\texttt{seen-2-after-1}       		& 2	         & \textsc{SyntCondPrefix}& \texttt{max} 	& \texttt{(= element 2)}	& 3 \\
\texttt{alternation-of-11-22 }      	& 3	         & \textsc{SyntCondPrefix}& \texttt{min} 	& \texttt{(= element EOF)}	& 4 \\

\bottomrule

\end{tabular}%
} 
\end{center}%
\caption{Preliminary evaluation of \sympa.}
\label{tab:eval}%
\end{table}%

%% file: related.tex
\section{Related work}

The problem of parallelizing recurrent programs dates back to seventies~\cite{kogge}.
Today, from purely mathematical solutions it grew up to the mature large-scale industrial applications within MapReduce~\cite{DBLP:conf/osdi/DeanG04}, Dryad~\cite{DBLP:conf/eurosys/IsardBYBF07}, Spark~\cite{DBLP:conf/nsdi/ZahariaCDDMMFSS12} and Hadoop~\cite{DBLP:books/daglib/0022835} distributed programming platforms.
To automatically generate MapReduce programs, \cite{DBLP:conf/oopsla/RadoiFRS14} proposes to translate sequential code into parallel one based on the set of rewrite rules.
Alternatively, the MapReduce program synthesis can be driven by input/output examples~\cite{DBLP:conf/pldi/SmithA16}.
Instead of synthesizing the MapReduce code, \cite{DBLP:conf/sosp/RaychevMM15} proposes to synthesize symbolic summaries over concrete segments of the input data, the combination of which in a predefined order produces the final output.
\sympa can be also viewed as a technique to synthesize a MapReduce-like code, and in contrast to the aforementioned works, it exploits template-based synthesis and bounded model checking.

Other approaches targeting synthesis of concurrent programs include~\cite{DBLP:conf/pldi/VechevY08,DBLP:conf/ppopp/BartheCKGM13,DBLP:conf/sc/XuKS14}.
In particular, \cite{DBLP:conf/sc/XuKS14} deals with partial SPMD programs which are automatically converted into sequential partial programs, for which the requested code fragments are synthesized.
The approach of~\cite{DBLP:conf/pldi/VechevY08} synthesizes synchronization primitives to be used in concurrent programs.
The approach of~\cite{DBLP:conf/ppopp/BartheCKGM13} performs automatic vectorization, i.e., replaces loop-free blocks of code from the loop bodies with sequences of synthesized SIMD instructions.
All these works use formal methods, but they are applicable to the different classes of problems than the one of \sympa.

Programming By Examples (PBE) is nowadays one of the most promising directions in synthesis~\cite{DBLP:journals/cacm/GulwaniHS12,DBLP:conf/pldi/BarowyGHZ15,DBLP:conf/pldi/FeserCD15,DBLP:conf/cav/AlurCR15,DBLP:conf/pldi/SmithA16}.
Since the applications of PBE do not require an explicit specification (contrary to \sympa that treats the sequential program as specification), PBE is orthogonal to our synthesis approach.
However, one can still find a connection between bounded synthesis and PBE.
Indeed, while fixing a bound for the number of elements in the given array, we force \sympa to implicitly consider all possible examples consisting of the bounded arrays and the corresponding outputs of the given sequential program.

Formal methods are getting more involved into the data-parallel and array-handling computing: \cite{DBLP:conf/cav/ChenSW16} studies the commutativity problem of MapReduce, \cite{DBLP:conf/cav/DacaHK16} proposes an approach to prove counting properties in unbounded array-handling programs.
Finally, equivalence checking is one of the most intriguing branches of formal methods (e.g.,~\cite{KK1997,DBLP:conf/oopsla/0001SCA13,DBLP:conf/dac/ClarkeKY03,os2009,DBLP:conf/cav/LahiriHKR12,pde}).
Currently, \sympa uses equivalence checking only in the bounded context, to verify that a candidate decomposition is equivalent to the given sequential program.
In future, it would be interesting to see which unbounded equivalence checking methods can also be applicable to \sympa.

\section{Conclusion}

In this paper, we addressed the challenge of bringing the existent sequential array-handling code into a parallel fashion.
We presented an approach called \sympa that applies for the cases when the data is partitioned into a sequence of segments, each segment processed separately, and the partial outputs for the segments are merged together.
\sympa augments the given sequential program with an additional functionality to decompose data dependencies in loop iterations.
\sympa gradually considers several parallelization scenarios and attempts to find easier solutions first.
We presented a prototype of the algorithm \sympa to automatically synthesize decompositions of small programs in Racket, and we envision its further improvements in future.

\paragraph{Acknowledgments.}
This work is supported in part by the SNSF Fellowship P2T1P2\_161971, NSF Grants CCF--1139138, CCF--1337415, and NSF ACI--1535191, a Grant from U.S. Department of Energy, Office of Science, Office of Basic Energy Sciences Energy Frontier Research Centers program under Award Number FOA--0000619, and grants from DARPA FA8750--14--C--0011 and DARPA FA8750--16--2--0032 , as well as gifts from Google, Intel, Mozilla, Nokia, and Qualcomm.